\documentclass{PoS}
\usepackage{epsf}

\newcommand{\half}{\mbox{\small $\frac{1}{2}$}}          
\newcommand{\third}{\mbox{\small $\frac{1}{3}$}}         
\newcommand{\ti}{\mbox{\tiny $T\!I$}}                    
\newcommand{\lat}{\mbox{\tiny $L\!A\!T$}}                
\newcommand{\WI}{\mbox{\tiny $WI$}}                      
\newcommand{\plaquette}{\mbox{\tiny $Plaquette$}}        
\newcommand{\rectangle}{\mbox{\tiny $Rectangle$}}        
\newcommand{\QCD}{\mbox{\tiny $Q\!C\!D$}}                

\def\lsim{\mathrel{\rlap{\lower4pt\hbox{\hskip1pt$\sim$}}
    \raise1pt\hbox{$<$}}}                
\def\gsim{\mathrel{\rlap{\lower4pt\hbox{\hskip1pt$\sim$}}
    \raise1pt\hbox{$>$}}}                
\title{
\vspace*{-2cm}
\begin{minipage}{\textwidth}
\begin{flushright}
\texttt{\footnotesize
PoS(LAT2008)132    \\%
DESY 08-164        \\%
Edinburgh 2008/31  \\%
Liverpool LTH 812  \\%
}
\end{flushright}
\end{minipage}\\[15pt]
\vspace*{+2cm}
       Clover improvement for stout-smeared 2+1 flavour 
       SLiNC fermions: non-perturbative results}

\ShortTitle{Non-perturbative determination of $c_{sw}$ for SLiNC fermions}

\author{N.~Cundy$^{a}$,
        M.~G\"ockeler$^{a}$,
        \speaker{R.~Horsley}$^{\,b}$,
        T.~Kaltenbrunner$^{a}$,
        A.~D. Kennedy$^{b}$,
        Y.~Nakamura$^c$,
        H.~Perlt$^{d}$,
        D.~Pleiter$^c$,
        P.~E.~L.~Rakow$^e$,
        A.~Sch\"afer$^{a}$,
        G.~Schierholz$^{ac}$,
        A.~Schiller$^{d}$,
        H.~St\"uben$^f$
        and J.~M.~Zanotti$^b$ \\
        \llap{$^a$} Institut f\"ur Theoretische Physik,
                    Universit\"at Regensburg,
                    93040 Regensburg, Germany \\
        \llap{$^b$} School of Physics and Astronomy,
                    University of Edinburgh,
                    Edinburgh EH9 3JZ, UK \\
        \llap{$^c$} John von Neumann Institute NIC / DESY Zeuthen,
                    15738 Zeuthen, Germany \\
        \llap{$^d$} Institut f\"ur Theoretische Physik,
                    Universit\"at Leipzig,
                    04109 Leipzig, Germany \\
        \llap{$^e$} Theoretical Physics Division,
                    Department of Mathematical Sciences,
                    University of Liverpool,
                    Liverpool L69 3BX, UK \\
        \llap{$^f$} Konrad-Zuse-Zentrum f\"ur Informationstechnik Berlin,
                    14195 Berlin, Germany \\
        E-mail: \email{rhorsley@ph.ed.ac.uk} }

\author{QCDSF--UKQCD Collaborations}

\abstract{We discuss an action in which the fermion matrix
has single level stout smearing for the hopping terms together
with unsmeared links for the clover term. With the (tree level)
Symanzik improved gluon action this constitutes the {\bf S}tout {\bf Li}nk
{\bf N}on-perturbative {\bf C}lover or SLiNC action.
To cancel $O(a)$ terms the clover coefficient, $c_{sw}$
has to be tuned. We present here preliminary results of
a non-perturbative determination of $c_{sw}$ using the Schr\"odinger
functional and as a by-product also a determination of the critical
hopping parameter, $\kappa_c$. A determination of the renormalisation
constant for the local vector current is also given. Comparisons of
the results are made with lowest order perturbation theory results.}

\FullConference{The XXVI International Symposium on Lattice Field Theory\\
		 July 14-19 July 2008\\
		 Williamsburg, USA}

\begin{document}


\section{$O(a)$ Improvement}
\label{improvement}


When constructing a lattice QCD action, even the simplest gluon Lagrangian
action has only $O(a^2)$ corrections. The naive fermion action also has
$O(a^2)$ corrections, but suffers from the `doubling problem' describing
$16$ flavours in the continuum limit. A `cure' is to add the Wilson mass term,
so $15$ flavours decouple in the continuum limit, but the price is that
there are now $O(a)$ corrections, so that for example for a ratio of
hadron masses
\begin{eqnarray}
   {m_H \over m_{H^\prime}} = r_0 + ar_1 + O(a^2) \,.
                                                \nonumber
\end{eqnarray}
The Symanzik approach is a systematic improvement to $O(a^n)$ 
(where in practice $n=2$) by adding a basis (an asymptotic
series) of irrelevant operators and tuning their coefficients to remove
completely $O(a^{n-1})$ effects. Restricting improvement to on-shell
quantities the equations of motion reduce the set of operators in
both the action and in matrix elements. Indeed, for $O(a)$ improvement
only one additional operator in the action is required
\begin{eqnarray}
  {\cal L}_{clover} \propto a c_{sw}
     \sum \overline{\psi} \sigma_{\mu\nu}F_{\mu\nu}\psi \,,
                                                \nonumber
\end{eqnarray}
the so-called  `clover term'. So if we can improve {\it one} on-shell
quantity this then fixes $c_{sw}$ as a function of the lattice spacing
$a$ or equivalently of $g_0^2$, so that all other physical on-shell
quantities are automatically improved to $O(a)$, i.e., we now have
\begin{eqnarray}
  {m_H \over m_{H^\prime}} = r_0 + O(a^2) \,.
                                                \nonumber
\end{eqnarray}
Matrix elements still require additional $O(a)$ operators, for example
\begin{eqnarray}
   {\cal A}_\mu &=& ( 1 + b_Aam_q )( A_\mu + c_A a\partial^{\lat}_\mu P )
                                           \nonumber \\
   {\cal P}     &=& ( 1 + b_Pam_q ) P \,,
                                           \nonumber
\end{eqnarray}
with
\begin{eqnarray}
   A_\mu = \overline{q}\gamma_\mu\gamma_5 q \,, \qquad 
   P     = \overline{q}\gamma_5 q \,.
                                           \nonumber
\end{eqnarray}
An easily determined quantity is the quark mass, determined from
the PCAC relation%
\footnote{This is equivalent to considering the renormalised quark mass 
$m^{\WI}_{qR} = {Z_A (1 + b_Aam_q) \over Z_P(1 + b_Pam_q)} \, m_q^{\WI}$
as the difference is just a numerical factor, which in the chiral
limit does not effect considerations of $O(a)$-improvement.}
\begin{eqnarray}
   m_q^{\WI} = { \langle 
                \partial^{\lat}_0 ( A_0(x_0) + c_A a\partial^{\lat}_0 P(x_0)) O
                 \rangle
                 \over 2 \langle P(x_0) O \rangle } \,.
                                           \nonumber
\end{eqnarray}
Choosing different boundary conditions or operators, $O$, gives different
determinations of the quark mass $m_{q}^{\WI\,(i)}$, $i = 1$, $2$.
If the quark mass is improved then its errors are $O(a^2)$.
So we can determine improvement coefficients, $c_{sw}, \ldots$, by
finding the point where
\begin{eqnarray}
   m_{q}^{\WI\,(1)} = m_{q}^{ \WI\,(2)} \,.
                                           \nonumber
\end{eqnarray}
The ALPHA Collaboration achieved this by means of the 
`Schr\"odinger functional', \cite{luscher92a}.
Dirichlet boundary conditions are applied
on the time boundaries to the fields. For the gluon fields fixing
them on $x_0 = 0$ and $T$ is then equivalent to a constant chromo-electric
background field (which means that simulations with $m_q \sim 0$
with no zero mode problems are possible), while the fixed quark fields
($ \rho, \overline{\rho}$) can be taken as sinks/sources to build
operators for correlation functions. For example here we can take
at the lower boundary $x_0 = 0$ ($i=1$) and upper boundary $x_0 = T$ ($i=2$)
\begin{eqnarray}
   O^{(i)} = \sum_{\vec{y},\vec{z}} \,
               \left( - {\delta \over \delta \rho^{(i)}(\vec{y})} \right)
               \gamma_5
               \left( {\delta \over \delta \overline{\rho}^{(i)}(\vec{z})}
               \right) \,.
                                           \nonumber
\end{eqnarray}
So we can investigate PCAC behaviour at different distances from the
boundaries. Redefining the quark mass slightly (but in a way which coincides
to $O(a^2)$ in the improved theory, \cite{luscher96a}) to eliminate the
unknown $c_A$ ($m_q^{\WI} \to M$) we can define improvement when
\begin{eqnarray}
   (M, \Delta M) = (0, 0) \,,
                                           \nonumber
\end{eqnarray}
where
\begin{eqnarray}
   M \equiv M^{(1)} \qquad  \Delta M \equiv M^{(1)} - M^{(2)} \,,
                                           \nonumber
\end{eqnarray}
are chosen at some suitable $x_0$, \cite{luscher96a}. This gives the
required critical $c_{sw}^*$ and $\kappa_c^*$.

There are (small) ambiguities due to the finite volume used.
In an infinite volume we expect $O(a \Lambda_{\QCD})$ contributions
(in the chiral limit, otherwise there are also extra $O(am_q)$ terms)
due to the different boundary conditions or operators chosen.
In a finite volume there are additional $O(a/L_s)$ terms.
$O(a \Lambda_{\QCD}) \to 0$ as $a$ (or $g_0^2$) $\to 0$, but
$O(a/L_s) \sim O(1/N_s)$ (where $L_s = aN_s$).
We can either keep $L_s$ fixed in physical units as $a \to 0$ (the
`constant physics condition') so $O(a/L_s) \to 0$, or alternatively
simulate for several values of $N_s$ and extrapolate to $N_s \to \infty$.
The `Poor man's solution' is to evaluate at large $\beta \to \infty$
(small $a$) and subtract this result. Practically we have found that
for $c_{sw}$ this $O(1/N_s)$ term is negligible, while for $Z_V$,
this subtraction is about a $1\%$ affect.


\section{The SLiNC action}


\noindent
We shall apply the Schr\"odinger functional formalism
to $2+1$ flavour stout link clover fermions -- SLiNC fermions
(Stout Link Non-perturbative Clover). In a little more detail
\begin{eqnarray}
   S_F &=& \sum_x \left\{
   \kappa \overline{\psi}(x)\tilde{U}_\mu(x+\hat{\mu})
                                    [\gamma_\mu - 1] \psi(x-\hat{\mu}) -
   \kappa \overline{\psi}(x)\tilde{U}^\dagger_\mu(x-\hat{\mu})
                                    [\gamma_\mu + 1] \psi(x+\hat{\mu})
                  \right.
                                                \nonumber              \\
       & & \hspace*{0.25in} \left.
            + \overline{\psi}(x)\psi(x) +
             \half c_{sw}(g_0^2)
                 \overline{\psi}(x)\sigma_{\mu\nu}F_{\mu\nu}(x)\psi(x)
                            \right\} \,.
                                                \nonumber
\end{eqnarray}
The hopping terms (Dirac kinetic term and Wilson mass term)
use a once iterated stout smeared link or `fat link',
\begin{eqnarray}
   \tilde{U}_\mu
            &=& \exp\{iQ_\mu(x)\}\, U_\mu(x)
                                                            \nonumber \\
   Q_\mu(x) &=& {\alpha \over 2i} \left[ VU^\dagger - UV^\dagger 
                         - \third \mbox{Tr} (VU^\dagger - UV^\dagger) \right] \,,
                                                           \nonumber
\end{eqnarray}
($V_\mu$ is the sum of all staples around $U_\mu$)
while the clover term remains built from `thin' links -- they are
already of length $4a$ and we want to avoid the fermion matrix 
becoming too extended. Smearing is thought to help
at present lattice spacings and the stout variation is analytic
which means that the derivative can be taken (so the HMC force is
well defined) and perturbative expansions are also possible,
\cite{perlt08a}.

To complete the action we also use the Symanzik tree--level gluon action
\begin{eqnarray}
   S_G = {6 \over g_0^2} \, \left\{
         c_0 \sum_{\plaquette} {1 \over 3} \mbox{Re\,Tr}
             ( 1 - U_{\plaquette}) +
         c_1 \sum_{\rectangle} {1 \over 3} \mbox{Re\,Tr}
             ( 1 - U_{\rectangle})
                            \right\} \,,
                                                           \nonumber
\end{eqnarray}
together with
\begin{eqnarray}
   c_0 = {20 \over 12} \,, \,\,\, c_1 = - {1 \over 12}
              \qquad \mbox{and} \qquad
   \beta = {6c_0\over g_0^2} = {10 \over g_0^2} \,.
                                                           \nonumber
\end{eqnarray}


\section{The lattice simulation}


The lattice simulation used the Chroma software library, \cite{edwards04a}.
The Schr\"odinger Functional details follow \cite{klassen97a}.
All results were generated on $8^3\times 16$ lattices using the
HMC algorithm. A mild smearing of $\alpha = 0.1$ was used.
A series of simulations were performed (typically generating
$O(3000)$ trajectories), quadratic and then linear interpolations of the
$M$, $\Delta M$ results being used to locate the critical point.

\begin{figure}[htb]
   \hspace{1.25in}
   \epsfxsize=9.00cm
      \epsfbox{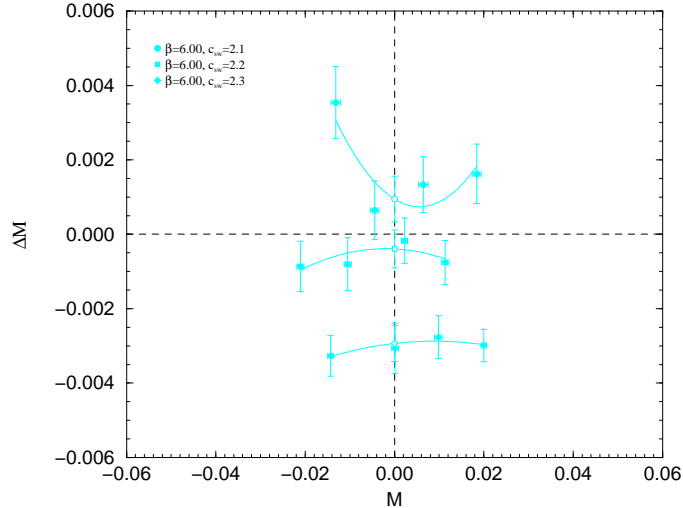}
   \caption{$\Delta M$ against $M$ for $\beta = 6.00$ (filled symbols)
            together with quadratic interpolations to $M = 0$
            (the open symbols).}
\label{M_dM}
\end{figure}

\begin{figure}[htb]
   \hspace{1.25in}
   \epsfxsize=9.00cm
\epsfbox{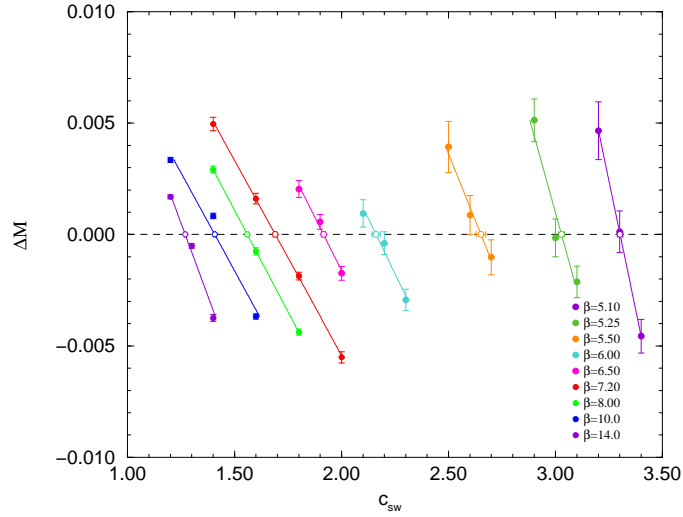}
   \caption{$\Delta M$ at $M = 0$ against $c_{sw}$ for various values
            of $\beta$ (filled circles) together with linear interpolations
            to $\Delta M = 0$ (open circles).}
\label{csw_dM}
\end{figure}

We thus have a two-parameter interpolation in $c_{sw}$ and $\kappa$
which is split here into two separate interpolations.
First plotting $\Delta M$ against $M$ and then interpolating to $M = 0$
gives $\Delta M(c_{sw},\kappa_c(c_{sw}))$. A typical result is shown in
Fig.~\ref{M_dM}.
These values of $\Delta M$ for $M = 0$ for various $\beta$ values are
then plotted against $c_{sw}$ as shown in Fig.~\ref{csw_dM}.
$\Delta M =0$ then gives $c_{sw}^*$.

A similar procedure yields $\kappa_c^*$: plotting $M$ against $1/\kappa$
and interpolating to $M = 0$ gives $\kappa_c(c_{sw})$. Then subsequently
plotting $\Delta M$ against $1/\kappa_c$ and interpolating to 
$\Delta M =0$ gives $\kappa_c^*$.


\section{Results}

\begin{figure}[htb]
   \hspace{1.25in}
   \epsfxsize=9.00cm
      \epsfbox{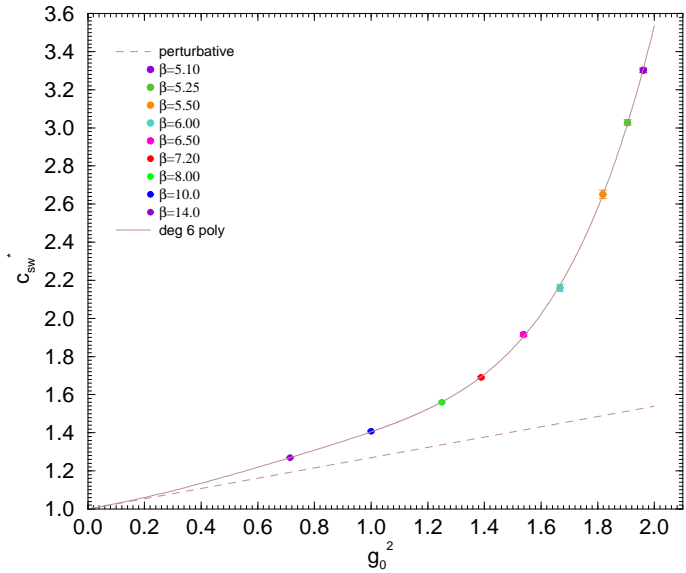}
   \caption{$c_{sw}^*$ against $g_0^2$ for various values of $\beta$
            (circles), together with a polynomial interpolation
            (line). Also shown is the perturbative result.}
\label{cswstar}
\end{figure}
\begin{figure}[htb]
   \hspace{1.25in}
   \epsfxsize=9.00cm
      \epsfbox{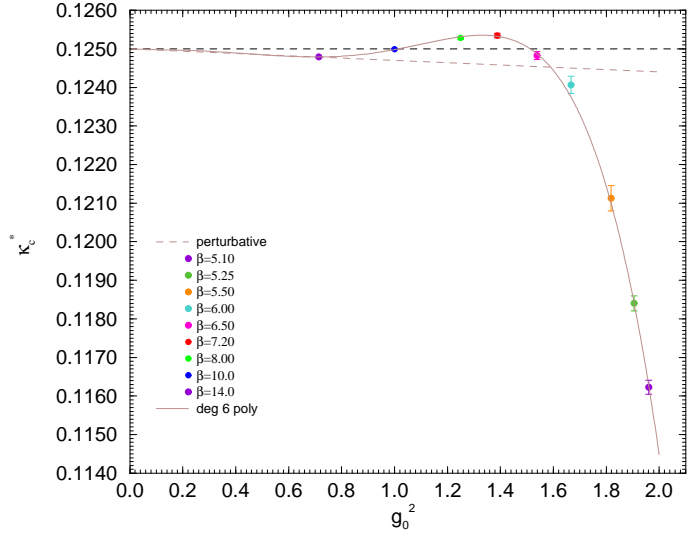}
   \caption{$\kappa_c^*$ against $g_0^2$ for various values of $\beta$
            (circles), together with a polynomial interpolation
            (line). Also shown is the perturbative result.}
\label{kapcstar}
\end{figure}

The results for $c_{sw}^*$ and $\kappa_c^*$ against $g_0^2$ are plotted
in Figs.~\ref{cswstar}, \ref{kapcstar} respectively in the range
$\beta \le 5.10$. The lowest order perturbative limit has been computed
for both $c_{sw}^*$ and $\kappa_c^*$, \cite{perlt08a} and is also shown
in the figures. An interpolation between the numerically determined
points is also shown. For both $c_{sw}^*$ and $\kappa_c^*$ a $6$th order
polynomial in $g_0^2$ proved sufficient. (These interpolation functions
are constrained to reproduce the perturbative results, \cite{perlt08a},
in the $\beta \to \infty$ limit. Therefore, they have four free parameters.)
This smooth fit between the points gives an estimate for $c_{sw}^*$
which will be used in the action for future generation of configurations.

For $c_{sw}^*$ the polynomial only tracks the perturbative solution
for small values of $g_0^2$. This is perhaps not surprising as
the tadpole improved, $TI$, estimate is $c_{sw}^{\ti} = u_0^{(S)}/u_0^4$,
\cite{perlt08a}, which is to be compared with the unsmeared case
of $c_{sw}^{\ti} = 1/u_0^3$ where $u_0$ is the average plaquette value
and $u_0^{(S)}$ is the smeared value. As smearing increases
the plaquette value this indicates that $c_{sw}^*$ can be large.
For $\kappa_c^*$ on the other hand as $\kappa_c^{\ti} = 1/(8u_0^{(S)})$
we expect that it is $\sim 1/8$. This is true for reasonably fine lattices,
however $\kappa_c^*$ does begin to decrease for larger values of $g_0^2$.
For $n_f=2$ the same phenomenon occurs: for larger $g_0^2$,
$\kappa_c^*$ begins to decrease (after initially increasing).

\begin{figure}[htb]
   \hspace{1.25in}
   \epsfxsize=9.00cm
      \epsfbox{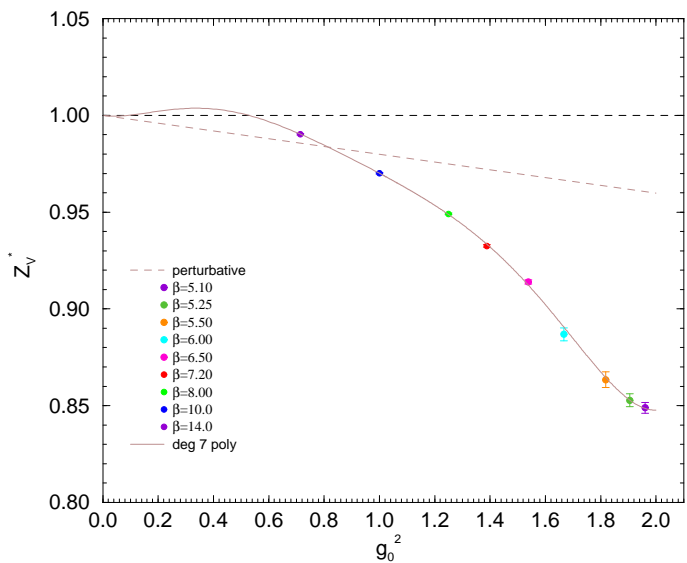}
   \caption{$Z_V^*$ against $g_0^2$ for various values of $\beta$
            (circles), together with a polynomial interpolation
            (line). Also shown is the perturbative result.}
\label{ZVstar_mod_g02_080919}
\end{figure}

Finally in Fig.~\ref{ZVstar_mod_g02_080919} we show the vector
renormalisation constant. This is computed using the vector current
in the ratio of a three-point to two-point function where the sinks/sources
are built using $O^{(i)}$ as described earlier in section~\ref{improvement}.


\section{Conclusions}


Non-perturbative $O(a)$ improvement is a viable procedure for (stout)
smeared actions with typical clover results being obtained.
(Other recent results for $2+1$ flavours are given in
\cite{yamada04a,edwards08a}.)
As $a$ decreases we need a significant $c_{sw} \gg c_{sw}^{tree} \equiv 1$
for $O(a)$ improvement. We are now seeking a region where
$a \sim 0.05 \, - \, 0.1\, \mbox{fm}$. Improvement, which is presumably
an asymptotic series, brings an advantage for smaller $a$ say
$a \le 0.1\,\mbox{fm}$. The two extremes for $a$ are simulations at
small $a$ with `large' $m_{ps}$ when there is no continuum
extrapolation but a chiral extrapolation, or alternatively
simulations at `coarse' $a$ with $m_{ps} \sim m_{\pi}$
when there is no chiral extrapolation but a continuum extrapolation.
Of course the Schr\"odinger functional does not tell us $a$;
for this conventional HMC simulations are required. Some preliminary results
indicate that around $\beta \lsim 6.0$ we have $a \lsim 0.07\,\mbox{fm}$.
Final results, including larger lattice size comparisons
will be published elsewhere, \cite{cundy08a}.


\section*{Acknowledgements}


The numerical calculations have been performed on the
BlueGeneLs at EPCC (Edinburgh, UK), NIC (J\"ulich, Germany),
the QCDOC (Edinburgh, UK) and the SGI ICE at HLRN
(Berlin-Hannover, Germany).
The BlueGene and QCDOC codes were optimised using Bagel, \cite{boyle05a}.
This work has been supported in part by
the EU Integrated Infrastructure Initiative Hadron Physics (I3HP) under
contract RII3-CT-2004-506078 and by the DFG under contracts
FOR 465 (Forschergruppe Gitter-Hadronen-Ph\"anomenologie) and
SFB/TR 55 (Hadron Physics from Lattice QCD).




\begin{thebibliography}{99}

\bibitem{luscher92a}
   M. L\"uscher \emph{et al.},
   \emph{Nucl. Phys.} {\bf B384}, 168 (1992)
   [{\tt arXiv:hep-lat/9207009}];
   \emph{Nucl. Phys.} {\bf B478}, 365 (1996)
   [{\tt arXiv:hep-lat/9605038}].

\bibitem{luscher96a}
   M. L\"uscher \emph{et al.},
   \emph{Nucl. Phys.} {\bf B491}, 323 (1997)
   [{\tt arXiv:hep-lat/9609035}].

\bibitem{perlt08a}
   R. Horsley \emph{et al.}, QCDSF Collaboration,
   \emph{Phys. Rev. } {\bf D78}, 054504 (2008) 
   [{\tt arXiv:0807.0345}];
   H. Perlt,
   QCDSF Collaboration, talk at Lattice 2008,
   arXiv:0809.4769.

\bibitem{edwards04a}
   R. Edwards and B. Jo{\'o},
   \emph{Nucl. Phys. Proc. Suppl.} {\bf 140}, 832 (2005)
   [{\tt arXiv:hep-lat/0409003}].

\bibitem{klassen97a}
   T. Klassen,
   \emph{Nucl. Phys.} {\bf B509}, 391 (1998)
   [{\tt arXiv:hep-lat/9705025}].

\bibitem{yamada04a}
   N.Yamada, \emph{et al.}, CP-PACS, JLQCD Collaborations,
   \emph{Phys. Rev.} {\bf D71}, 054505 (2005)
   [{\tt arXiv:hep-lat/0406028}];
   S. Aoki, \emph{et al.}, CP-PACS, JLQCD Collaborations,
   \emph{Phys. Rev.} {\bf D73}, 034501 (2006)
   [{\tt arXiv:hep-lat/0508031}].

\bibitem{edwards08a}
   R.~G. Edwards \emph{et al.},
   arXiv:0803.3960.

\bibitem{cundy08a}
   N. Cundy \emph{et al.}, QCDSF--UKQCD Collaborations,
   in preparation.

\bibitem{boyle05a}
   P.~A. Boyle,
   http://www.ph.ed.ac.uk/$\sim$paboyle/Bagel.html (2005).

\end{thebibliography}
\end{document}